\documentstyle[aps,amsmath,latexsym,amssymb,epsfig,multicol,graphicx]{revtex}

\begin{document}

\draft
\pagestyle{myheadings}

\newcommand{\comments}[1]{\hfill {\tt {eq:~#1}}} 

\def\breakon{\end{multicols}\widetext\vspace{-.2cm}
\noindent\rule{.48\linewidth}{.3mm}\rule{.3mm}{.3cm}\vspace{.0cm}}

\def\breakoff{\vspace{-.2cm}
\noindent
\rule{.52\linewidth}{.0mm}\rule[-.27cm]{.3mm}{.3cm}\rule{.48\linewidth}{.3mm}
\vspace{-.3cm}
\begin{multicols}{2}
\narrowtext}

\title{Polariton condensation and lasing in
optical microcavities - the decoherence driven crossover}  

\author{M. H. Szymanska, P. B. Littlewood, and  B. D. Simons}
\address{ Theory of Condensed Matter, Cavendish Laboratory, Cambridge
CB3 0HE, UK }

\date{\today}

\maketitle

\begin{abstract}
We explore the behaviour of a system which consists of a photon mode
dipole coupled to a medium of two-level oscillators in a microcavity
in the presence of decoherence. We consider two types of
decoherence processes which are analogous to magnetic and non-magnetic
impurities in superconductors. We study different phases of this
system as the decoherence strength and the excitation density is
changed. For a low decoherence we obtain a polariton condensate with
comparable excitonic and photonic parts at low densities and a
BCS-like state with bigger photon component due to the fermionic phase
space filling effect at high densities. In both cases there is a large
gap in the density of states. As the decoherence is increased the gap
is broadened and suppressed, resulting in a gapless condensate and
finally a suppression of the coherence in a low density regime and a
laser at high density limit. A crossover between these regimes is
studied in a self-consistent way analogous to the Abrikosov and Gor'kov
theory of gapless superconductivity \cite{abrikosov-gorkov}.

\end{abstract}

\begin{multicols}{2}
\narrowtext

\section{Introduction}

The miniaturisation and improvement in the quality of optical cavities
in recent years led to the achievement of strong-coupling regime of
light-matter interaction in many physical systems.  The strong
coupling regime is characterised by well-developed coupled modes of
light and electronic excitations, called polaritons.  Polariton
splitting has been experimentally observed for atoms \cite{atompol},
quantum wells \cite{cavpol} and bulk excitons \cite{bulkcavpol},
excitons in organic semiconductors \cite{organicpol1,organicpol2},
exciton complexes \cite{chargedpol} and glass spheres. The
strong coupling regime has also been achieved for coupled Josephson
junctions \cite{JJ} in a microwave cavity.

With well-developed modes, that are sharp and have long lifetimes, a
natural question becomes the existence of coherent, condensed states.
A theoretically constructed polariton condensate is a mixture of
coherent state of light and coherent state of massive particles in the
media. It is characterised by two order parameters: the coherent
polarisation and the coherent photon field and exhibit a gap in the
excitation spectrum \cite{paul}. Since the polariton condensate would
be a source of coherent light the natural question arises how it is
different from and how it can be connected to the traditional laser.

The laser is a weak-coupling phenomenon: a coherent state of photons
created by stimulated emission from an inverted electronic population
due to strong pumping. The polarisation of the medium is heavily
damped and the atomic coherence is practically zero.  A coherent
photon field, oscillating at the bare cavity mode frequency, is the
only order parameter in the system \cite{laser}.

The crossover between a polariton condensate and a laser is sometimes
mistakenly attributed to an increase in density and a crossover from
bosonic (exciton) degrees of freedom to fermionic particle-hole
pairs. The absence of polariton splitting is associated with
disappearance of coherence, which is why experiments are concentrated
at low densities. In this work we show that the trend is an opposite
one - the condensate is more robust at high densities. One needs to
remember that polaritons, and so polariton splitting, are normal state
excitation and so disappearance of polariton splitting in the normal
state does not indicate what would happen if the system was condensed.

The issue of electronic coherence in polariton condensate is
independent of whether the excitations are more ``excitonic'' or more
like a two-component plasma (ionised electron-hole pairs), or indeed
whether or not there is saturation (phase-space filling) of the
electronic states.  It has been shown \cite{paul} that although the
increase in the density of electronic excitation leads to
nonlinearities in the polariton system which cause the collapse of the
splitting between the two polariton peaks in the normal state it does
not destroy condensation even at very high excitation densities.  The
saturation in the fermionic space forces the condensate to become more
photon like as the excitation density is increased and to have more
BCS-like character, but nevertheless the coherence in the media and
the gap in the excitation spectrum are present.  The change in the
density of the electronic excitations leads to the crossover between
the regime of BEC of polaritons and collective BCS-like state

This article will argue that the real enemy of condensation is
decoherence, not density, and that it is precisely decoherence which
drives the polariton condensate towards the laser regime. We show that
only the self-consistent inclusion of decoherence processes allows to
establish a crossover between an isolated condensate and a laser. The
widely used quantum Maxwell-Bloch (Langevin) equations with constant
decay rate for polarisation are not correct in a regime when the
coherent polarisation is large and the gap in the density of states is
present. We develop a self-consistent method analogous to the
Abrikosov and Gor'kov theory of gapless superconductors, which allows
us to study the stability of the polariton condensate at low
decoherence strength and to established the crossover to laser
behaviour as the decoherence is increased.

\section{Model}
\label{model}

The model we consider in this work is schematically shown in Figure
\ref{fig:model}.  It consists of a set of N two-level oscillators
dipole coupled to a single mode of electromagnetic field confined in a
cavity. This system is then subject to various decoherence, pumping
and damping processes.  These processes can be of a different physical
nature, depending on the material, but their exact details are not
that important for a general theory. They can be described, similarly
to laser theory, as baths of harmonic oscillators coupled to the
system in a way that gives the same effect as the real physical
environment.

\breakon
\begin{figure}[htbp]
	\begin{center}
\includegraphics[width=16.2cm,angle=0]{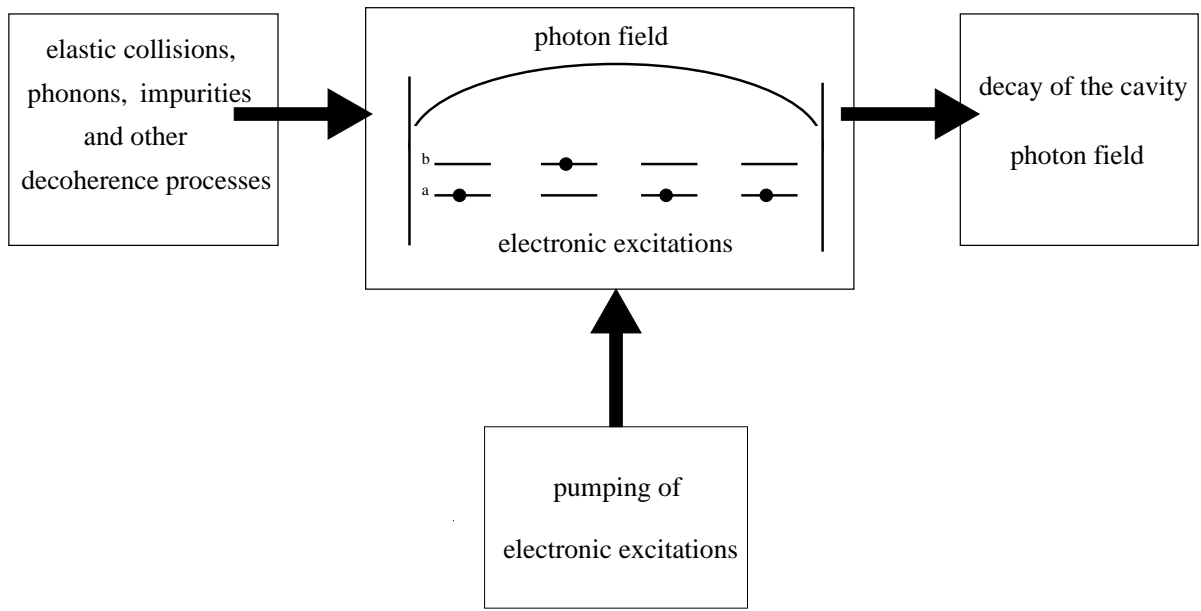} 
	\end{center}
	\caption{Sketch of the model studied in this work: the system
	of two-level oscillators dipole coupled to a single cavity
	mode interacting with various types of environment.}
\label{fig:model}
\end{figure}
\breakoff

Our model takes into account the major Coulomb interaction between the
electron and hole within the exciton, the phase-space filling effect,
disorder in the material (inhomogeneous broadening of excitonic
energies) and various types of decoherence effects.  However, it does
not include screening and Coulomb interactions between
excitons. Therefore, it gives a very good description of tightly
bound, Frenkel - type of excitons localised by disorder or bound on
impurities, molecular excitons in organic materials, atoms in the
solid state or Josephson junctions arrays in microwave
microcavity. And it gives only a qualitative description within a
mean-field approximation for other types of excitons like Wannier
excitons or excitons propagating in a sample (see Section
\ref{aplic}).

The model we use is the minimal required to describe the essential
physics. At this stage we do not intend to model any particular medium
with its complex interactions which would only make the general
picture less clear. More specific details of the particular medium
could, however, be included straightforwardly into the formalism. We 
will discuss this possibility in the Section~\ref{aplic}.

In the following, we consider a system comprised of an ensemble of $N$ 
two-level oscillators with an energy $\epsilon_j$ dipole coupled to a 
single cavity mode. The corresponding microscopic Hamiltonian takes the 
form
\begin{eqnarray}
&&\hat{H}_S = \omega_c\psi^\dagger\psi + \sum_{j=1}^N \epsilon_j
(b^\dagger_jb_j-a^\dagger_ja_j)\nonumber\\ 
&&\qquad\qquad + \sum_{j=1}^N \frac{g_j}{\sqrt{N}}(b^\dagger_ja_j\psi 
+\psi^\dagger a^\dagger_jb_j)
\label{HS}
\end{eqnarray}
where Fermionic operators $b_j$ and $a_j$ annihilate electrons in the upper 
and lower states respectively, while the Bosonic operator $\psi$ annihilates 
the photon. Here the sum extends over the possible sites $j$ where an 
exciton can be present (e.g. different molecules or localised states 
associated with a disorder potential). Matrix elements $g_j/\sqrt{N}$ 
describe the interaction of the photon with the two-level oscillators. 

Generally, the effect of the environment on the behaviour of the system 
can be modelled through the interaction of the internal degrees of freedom
with a bath. Taking into account different physical processes, the most
general coupling is of the form
\begin{eqnarray}
&&\hat{H}_{SB}=\sum_{k} g_{k}(\psi^\dagger d_{k}+d_{k}^\dagger \psi)
\nonumber\\
&&\qquad\qquad + \sum_{jk} \left[b^\dagger_j a_j 
(g^{\gamma_+}_{jk} c_{+,k}^\dagger+
g^{\gamma_-}_{jk} c_{-,k})+{\rm h.c}\right]\nonumber\\
&&\qquad\qquad + \sum_{jk} \Gamma^{(1)}_{jk}(b^\dagger_jb_j
+a^\dagger_ja_j)(c_{1,k}^\dagger +c_{1,k})\nonumber\\
&&\qquad\qquad + \sum_{jk} \Gamma^{(2)}_{jk}(b^\dagger_jb_j
-a^\dagger_ja_j)(c_{2,k}^\dagger +c_{2,k}),
\label{eq:HSB}
\end{eqnarray}
where 
\begin{eqnarray*} 
&&\hat{H}_{B}= \sum_{k} \left[\omega_{k} d_k^{\dagger}d_k +\omega_{+,k}
c^\dagger_{+,k} c_{+,k}+\omega_{-,k} c^\dagger_{-,k} c_{-,k}\right.\\
&&\qquad\qquad\qquad\qquad \left.+ \omega_{1,k} c^\dagger_{1,k} c_{1,k} 
+\omega_{2,k}c^\dagger_{2,k} c_{2,k}\right]
\end{eqnarray*}
describes the Hamiltonian of the bath. Here the different modes $k$ of the 
bath are indexed by (independent) Bosonic field operators $d_k$, $c_{+,k}$, 
$c_{-,k}$, $c_{1,k}$ and $c_{2,k}$. Here the first term in (\ref{eq:HSB}) 
describes the decay of the photon field from the cavity. Matrix elements
$g^{\gamma_+}_{jk}$ describe the incoherent pumping of two-level 
oscillators, while the matrix elements $g^{\gamma_-}_{jk}$ contain all
of the higher energy processes which destroy the electronic excitations 
such as the radiative decay
into photon modes different from the cavity mode. Apart from their dephasing
effect, together, these processes cause a flow of energy through the system. 
However, in the steady state, the total number of excitations
\begin{equation} 
\label{constrain}
\hat{n}_{ex}=\psi^\dagger\psi + \frac{1}{2}\sum_{j}(b^\dagger_jb_j-
a^\dagger_ja_j), 
\end{equation} 
the sum of photons and excited two-level oscillators, is constant. Finally, 
the third and fourth terms describe all those lower energy dephasing 
processes, such as
collisions and interactions with phonons and impurities, which conserve
the total number of excitations in the cavity. Such contributions can be 
divided into a part which act symmetrically ($\Gamma^{(1)}$) or 
antisymmetrically ($\Gamma^{(2)}$) on the upper and lower levels. Altogether 
these four terms contain all the essential mechanisms of decoherence.

To assimilate the effect of the different mechanisms of decoherence, we
will find it useful both intuitivelty as well as techincally (see later) 
to draw on an analogy between the Hamiltonian of the system and that of a
superconductor~\cite{keldysh1,keldysh2}. Referring to 
the states indexed by the Fermionic operators $b_j$ as `particle-like',
and those indexed by $a_j$ as `hole-like', $\hat{H}_S$ can be interpreted
as a BCS Hamiltonian for a superconductor with an imposed homogeneous 
`superconducting' order parameter $\psi$. With this analogy, it is clear
that the second and third terms of $\hat{H}_{SB}$~(\ref{eq:HSB}) affect 
a mechanism of `pair-breaking', while the fourth term is compatibile with
the symmetry of the Cooper pairs. The former act on the system as 
dynamically fluctuating magnetic impurities while the latter describe
dynamical fluctuations of a normal non-magnetic potential providing only 
an inhomogeneous broadening of the energies.

When the pumping and photon decay rates are high, the system would be 
driven out of equilibrium. However, if the thermalisation rate is in 
excess of the speed at which the system is pumped, an equilibrium 
assumption can be justified. In this work we will limit our considerations
to this regime focussing on the effect of decoherence on the equilibrium
system. Choosing both the pumping and decay rates to be small --- allowing
thermal equilibration --- their ratio can be used to fix the total 
excitation density $n_{ex}$. Nevertheless, within this quasi-equilibrium 
regime, both the third and fourth terms in (\ref{eq:HSB}) can, in fact, be 
arbitrary large since they do not couple to the total density $n_{ex}$. 
There is no restriction on the density of excitations nor on the decoherence 
rate.

Therefore, on this background, we will consider the total Hamiltonian
\begin{equation} 
\hat{H}=\hat{H}_S+\hat{H}_{SB}+\hat{H}_{B},
\label{H}
\end{equation} 
where
\begin{eqnarray}
&&\hat{H}_{SB} = \sum_{jk} \Gamma^{(1)}_{jk}(b^\dagger_jb_j
+a^\dagger_ja_j)(c_{1,k}^\dagger +c_{1,k})\nonumber\\
&&\qquad\qquad+\sum_{jk} \Gamma^{(2)}_{jk}(b^\dagger_jb_j
-a^\dagger_ja_j)(c_{2,k}^\dagger +c_{2,k}),
\label{Hmag}
\end{eqnarray}
includes only the includes only the decoherence mechanisms which conserve
$n_{ex}$, while 
\begin{eqnarray*} 
\hat{H}_{B}= \sum_{i=1,2} \sum_{k} \omega_{i,k} c^\dagger_{ik} c_{ik}.
\end{eqnarray*}
Although, in general, the coupling constants $\Gamma^{(i)}_{jk}$ can be site
dependent, for simplicity, we will suppose that the modes of the bath couple 
with equal strength to the two-level systems setting $\Gamma^{(i)}_{jk} 
\mapsto \Gamma^{(i)}_k$. Similarly, in the following, we will assume that the 
coupling of the two-level systems to the cavity photon is independent of 
the site index $j$, $g_j\mapsto g$.

When the interactions between the environment and the system are large, the
Hamiltonian above provides the basis of the standard theory of lasers (see,
e.g., Refs.~\cite{laser,haken}). At the same time, if we set the coupling
constants between the system and the environment to zero, the ground state
of the Hamiltonian $\hat{H}_S$ forms a polariton condensate. Thus by varying 
the magnitude of the coupling between the system and the baths the model
provides the means to move smoothly between an isolated condensate and
other phases driven by the decoherence. This provides a means to explore
the stability of the polariton condensate to interactions with the outside 
world at a small coupling strength, and to establish the connection between 
polariton condensation and lasers as the decoherence is increased.

The standard assumption (which in many cases is physically correct) is
that the environment leads to rapid dephasing of the exciton polarisation 
(i.e. $T_2$ is very short). It is thus generally assumed that the 
polarisation is very small, and that the coherent photon field is the
dominant order parameter. In such a case the baths can be averaged out 
before the exciton-photon interaction is studied. This in turn leads to the 
well-known quantum Maxwell-Bloch (Langevin) equations for the photon 
field and polarisation, essentially of the form
\begin{equation} 
\label{maxwell-bloch} \frac{d}{dt}
\langle a^\dagger b\rangle = i\langle[\hat{H}_{S}, a^\dagger b]\rangle
- \frac{1}{T_2} \langle a^\dagger b \rangle \;\;\;.  
\end{equation}
Crucially, Eq.~(\ref{maxwell-bloch}) can be derived by assuming that the
lifetime $T_2$ for polarisation of a \emph{single} two-level oscillator is
the same as that of the macroscopic ensemble of two-level oscillators (which
would correspond to introducing a separate bath for each two-level
system). However, in general Eq.~(\ref{maxwell-bloch}) is not correct
\cite{us0}. When $T_2$ is long, the macroscopic ensemble of two-level
oscillators can exist in a collective state characterised by a large
coherent polarisation, and the assumptions which lead to
(\ref{maxwell-bloch}) cannot be justified. Moreover, the constant
decay rate $1/T_2$ in (\ref{maxwell-bloch}) is a critical parameter;
even at arbitrarily small decoherence it leads to completely different
solutions from those in the absence of an environment (as shown in
Ref.~\cite{us0}). This criticality is however unphysical and arises only due
to approximations used in deriving the Langevin equations.
In fact, this conclusion seems not to be widely appreciated. Indeed, in a
relatively recent publication~\cite{wrong} similar decay constants in
the equations of motion for the coherently driven excitonic insulator
have been used, leading to the conclusion that the excitonic insulator
phase cannot exist for an arbitrary small decoherence.

One can gain some physical insight into this problem by considering the 
evolution of the density of states: The ideal condensate has a gap in 
the density of states which would still be present for small decoherence. 
It is evident that the coherent fields in this regime cannot be damped 
just by constant decay rates independent of frequency as there are no 
available states in which to decay. As the decoherence is increased this 
gap gets smaller and finally is completely suppressed causing the coherent 
fields to be strongly damped, as in lasers. In this regime 
Eq.~(\ref{maxwell-bloch}) is perfectly valid. However to be able to study 
a crossover from the fully phase coherent polariton condensate to a laser 
one needs to include the environment in a self-consistent way in which the 
possible gap in excitation spectrum and a large polarisation would be taken 
into account.

The analogy between superconductivity and the excitonic insulator was
first noticed and exploited in a pioneering work of Keldysh
\cite{keldysh1,keldysh2}, and explored later by others
\cite{zittart,paul}. The analogy between different types of
decoherence processes acting on polariton condensate, and the problem of
magnetic and potential impurities in superconductors suggests that similar
methods to that used by Abrikosov and Gor'kov (AG) in their theory of
gapless superconductivity~\cite{abrikosov-gorkov} can be useful in
studying the properties of polariton condensates in the presence of
decoherence. This analogy is however not exact and we will discuss a
few essential differences between our theory and the Abrikosov and Gor'kov
approach at the end of Section \ref{method}. An outline of the rest of
the paper is as follows: In section \ref{method} we provide details of
the method in the path integral formulation. In section \ref{gs} we
discuss ground state properties of the system in the presence of two
different types of dephasing while in section \ref{es} we study the
excitation spectrum and construct a phase diagram as a function of
excitation density, pair-breaking decoherence strength and the
inhomogeneous broadening of energy levels. In section \ref{cl} we
discuss a crossover between an isolated condensate and a laser and in
section \ref{gap} the magnitude of the energy gap. In section
\ref{aplic} we comment on the applicability of the model and the
method presented in this work and indicate a few directions in which
the model could be easily extended. Finally, in section \ref{exp} we
discuss recent experiments in the light of theoretical predictions
presented in this work and in section \ref{sum} we briefly summarise
the results.

\section{Path Integral Formulation}
\label{method}

To construct a theory of the coupled system, we will exploit an approach
based on the coherent state path integral. As well as providing 
access to the mean-field equations of the system, such a framework provides
the potential to explore the influence of fluctuation phenomena. Working
in the grand canonical ensemble, a chemical potential $\mu$ can be used
to fix the total number of excitations $n_{ex}$. The quantum partition 
function of the system ${\cal Z}={\rm Tr}\  e^{-\beta(\hat{H}-\mu
\hat{n}_{ex})}$ can be expressed as a coherent state path integral over 
Fermionic and Bosonic fields, 
\begin{equation} 
{\cal Z}=\int D[\bar{\psi}, \psi] \prod_j
D[\bar{\phi}_j,\phi_j] \prod_{k,i=1,2} D[\bar{c}_{i,k}, c_{i,k}] e^{-S}.
\label{partition}
\end{equation}
As with the Hamiltonian, it is convenient to separate the total action
as $S=S_S+S_{SB}+S_{B}$ where
\begin{eqnarray*}
S_S=S_\psi-\int_0^\beta d\tau \sum_j \bar\phi_j \hat{G}_{0j}^{-1} \phi_j
\end{eqnarray*}
with $S_\psi=\int_0^\beta d\tau \bar\psi(\partial_\tau+\omega_c-\mu)\psi$
denotes the action of the internal electron and photon degrees of freedom
of the system, while the action for the coupling of the system to the bath 
takes the form
\begin{eqnarray*}
&&S_{SB}=\int_0^\beta d\tau \sum_{i=1,2}\left[\sum_k \bar{c}_{ik}
(\partial_\tau +\omega_{ik}) c_{ik}\right.\\
&&\qquad\qquad\qquad\left.+\sum_k
\Gamma_k^{(i)}\rho^{(i)}(\bar{c}_{i,k}+ c_{i,k})\right].
\end{eqnarray*}
Combining the Fermionic fields of the two-level systems into a Nambu-like
spinor,
\begin{displaymath}
\bar{\phi}_j=\left( \begin{array}{c}
\bar{b}_{j}  \  \ \bar{a}_{j}
\end{array}
\right), 
\end{displaymath} 
the bare Green function assumes the matrix form
\begin{eqnarray*}
\hat{G}_{0j}^{-1}=-\partial_\tau\sigma_0-(\epsilon_j-\mu/2)\sigma_3-g\bar\psi
\sigma_- -g\psi\sigma_+\,.
\end{eqnarray*}
where $\sigma$ denote the Pauli spin matrices (with $\sigma_0\equiv\openone$)
which operate in the $(b,a)$ space (hereafter, loosely referred to as the 
`particle/hole' space). 
Finally, we have defined the symmetric and antisymmetric `densities' according 
to the relation $\rho(\tau)^{(1,2)}=\sum_j \bar{\phi}_j(\tau)\sigma_{(0,3)}
\phi_j(\tau)$.

Although a theory of the symmetric and antisymmetric processes can be 
developed in concert, for clarity we will present a detailed derivation
of the action of the `pair-breaking' decoherence processes imposed by 
$\Gamma^{(1)}$. Later, in Section \ref{saddle}, we will restore the 
decoherence processes affected by $\Gamma^{(2)}$. Thus, for now, we will 
use the following abbreviation $\Gamma_k^{(1)}\mapsto \Gamma_k$ dropping
the `channel' index. 

Being Gaussian in the Bosonic fields $c_{k}$, the degrees of freedom of the
bath can be integrated out leading to an effective interaction of the 
two-level systems which takes the form $\int D[\bar{c}_k,c_k] e^{-S_{SB}}
=e^{-S_{SB}'}$, where
\begin{equation}
\label{Sef}
S_{SB}'=\int_0^\beta d\tau d\tau' \rho(\tau) 
\sum_k \Gamma_k^2 D_k(\tau-\tau') \rho(\tau'),
\end{equation}
with $\hat{D}^{-1}_k=-\partial_\tau-\omega_k$ representing the free propagator 
of the environment. Transforming (\ref{Sef}) to the Fourier Matsubara 
frequency representation, and summing over the internal degrees of freedom 
of the bath,
\begin{equation}
\label{defgam}
-\sum_k \Gamma_k^2 D_k(i\nu_n)=f_{\Gamma}(i\nu_n),
\end{equation}
where $D^{-1}_k(\nu_n)=i\nu_n-\omega_k$ and $\nu_n=2\pi n/\beta$, the 
induced interaction assumes the form
\begin{eqnarray*}
&&S_{SB}'=-\sum_{\nu_n} f_{\Gamma}(i\nu_n) \rho(i\nu_n)\rho(-i\nu_n)\\
&&=\int_0^\beta d\tau d\tau' f_{\Gamma}(\tau-\tau') \sum_{jj'} {\rm Tr}
\, \phi_j(\tau)\otimes\bar\phi_{j'}(\tau') \phi_{j'}(\tau') \otimes 
\bar\phi_j(\tau).
\end{eqnarray*}
In particular, it can be seen explicitly that the interaction with the 
environment introduces an effective quartic interaction between the
\emph{different} two-level systems. This contrasts with the Maxwell-Bloch
equations~(\ref{maxwell-bloch}) from which one can infer only a lifetime 
for excitations~\cite{us0}.

To develop a mean-field theory of the coupled system, it is helpful to
affect a Hubbard-Stratonovich decoupling of the interaction. Introducing 
the $2\times 2$ component matrix field $Q_{jj'}(\tau,\tau')$, which 
inherits the symmetry of the dyadic product $\phi_j(\tau)\otimes
\phi_{j'}(\tau')$, the interaction generated by the bath can be decoupled
as 
\begin{eqnarray*}
&&e^{-S_{SB}'}=\int DQ e^{-S_Q}\\
&&\times\exp\left[\int_0^\beta d\tau d\tau' f_\Gamma (\tau-\tau')
\sum_{jj'} \bar{\phi}_j(\tau) Q_{jj'}(\tau,\tau')\phi_{j'}(\tau')
\right],
\end{eqnarray*}
where
\begin{eqnarray*}
S_Q=\int_0^\beta d\tau d\tau' \sum_{jj'} f_\Gamma (\tau-\tau')
{\rm Tr}\ Q_{jj'}(\tau,\tau') Q_{j'j}(\tau',\tau)
\end{eqnarray*}
with trace taken in the particle-hole space. Combined with $S_S$, an
integration over the Fermionic degrees of freedom $\phi$ obtains the 
quantum partition function
\begin{equation} 
{\cal Z}=\int D[\bar{\psi}, \psi] \int DQ e^{-S}.
\end{equation}
where $S=S_Q+S_\psi-{\rm Tr}\ln \hat{G}^{-1}$, 
\begin{equation}
\label{Gtot}
G^{-1}_{jj'}(\tau,\tau')=G_{0j}^{-1}\delta_{jj'}\delta(\tau-\tau')-
f_\Gamma(\tau-\tau') Q_{jj'}(\tau,\tau').
\end{equation}
and the trace now runs over time and site indicies as well as the 
particle/hole space. At this level, the analysis is exact. 

\subsection{Saddle-point equations}
\label{saddle}

To develop the quantum partition function, further progress is possible 
only within a saddle-point approximation. Varying the action with respect
to $Q$, one finds that each electronic excitation is coupled to the average 
field created by all of the other excitations. In this sense, the 
saddle-point analysis corresponds to a mean-field treatment of the 
interaction between electronic excitations. This analysis becomes exact 
when there is a large number of electronic excitations coupled to a small 
number of field modes, since the fluctuations of the field are then 
negligible~\cite{paul}. The mean-field treatment of the interaction 
becomes exact in the thermodynamic limit when $N\to\infty$. The fluctuations 
above mean-field are of the order of $1/\sqrt{N}$. 

In the saddle-point approximation, it is assumed that the dominant
contribution to the quantum partition function (\ref{partition}) arises from
those configurations $\psi$ and $Q$ which minimise the total action.
Varying the action $S$ with respect to $Q$ one obtains the matrix equation
\begin{equation} 
Q_{jj'}(\tau,\tau')=\frac{1}{2} G_{jj'}(\tau,\tau'),
\label{saddlepointQ}
\end{equation}
while varying with respect to $\bar{\psi}$, the saddle-point equation 
takes the form
\begin{equation} 
(\partial_\tau+\omega_c-\mu)\psi(\tau)=\sum_j g {\rm Tr}\ G_{jj}(\tau,
\tau)\sigma_-.
\label{saddlepointF}
\end{equation}
Maintaining the analogy with the superconductor, the first equation 
which identifies $Q$ as the self-energy, describes the self-consistent 
Born approximation for the Green function, while the 
second equation represents the gap equation for the superconducting order
parameter. Substituting Eq.~(\ref{saddlepointQ}) into (\ref{Gtot}) the
saddle-point equation assumes the form of a Dyson equation
\begin{equation}
\label{Dyson}
G^{-1}_{jj'}(\tau,\tau')=G_{0j}^{-1}(\tau,\tau')\delta_{jj'}-
\Sigma_{jj'}(\tau,\tau'), 
\end{equation}
where 
\begin{eqnarray*}
\Sigma_{jj'}(\tau,\tau') =
\frac{1}{2}f_{\Gamma}(\tau-\tau')G_{jj'}(\tau,\tau')
\end{eqnarray*}
denotes the self-energy

Until now, we have focussed on the impact of the `pair-breaking' perturbation
affected by the matrix elements $\Gamma^{(1)}$. Consideration of the 
symmetric perturbation $\Gamma^{(2)}$ follows straightforwardly. In doing
so, it may be confirmed that the structure of the saddle-point equations
are maintained while the self-energy takes the form
\begin{eqnarray*}
\Sigma_{jj'}(\tau,\tau') =
\frac{1}{2}f_{\Gamma}(\tau-\tau')\sigma_3 G_{jj'}(\tau,\tau')\sigma_3
\end{eqnarray*}

In steady state one expects the solution of the saddle-point equations to
depend only on $\tau-\tau'$. In this case, a transformation to Matsubara
frequencies leads to the relation
\begin{equation}
\Sigma(i\nu_n) =
\frac{1}{2}\sum_{\nu_n'}f_{\Gamma}(i\nu_n')G(i\nu_n-i\nu_n').  
\label{selfen1}
\end{equation}
Generally, the solution of the saddle-point equation depends sensitively 
on the particular spectrum of decoherence $f_\Gamma$ and must be determined
self-consistently. However, an explicit solution to the saddle-point equation
can be established in various limits.

Generally, in order to determine $f_{\Gamma}$ from Eq.~(\ref{defgam}) one
can assume that coupling constants of the system to the bath $\Gamma_{k}$, 
as well as the bath density of states $N(\omega_{k})$, are continuous functions
of frequency. In this case, one can replace the summation over $k$ with an
integral over $\omega_{k}$ ($\Gamma_{k} \mapsto \Gamma(\omega_{k})
$, $\sum_{k} \mapsto \int d\omega_k N_k(\omega_k)$) whereupon
\begin{eqnarray*}
f_\Gamma(i\nu_n) = \int d\omega_k \frac{N_k(\omega_k)
\Gamma^2(\omega_k)}{-i \nu_n-\omega_k}.
\end{eqnarray*}
Now, in general $f_\Gamma$ will exhibit a particular frequency dependence 
determined by the density of states $N(\omega_k)$ of the bath and coupling
constant $\Gamma(\omega_k)$. However two special cases present themselves:
Firstly, if we assume that both $\Gamma(\omega_k)$ and $N(\omega_k)$ are 
largely independent of frequency over a wide range, one finds 
$f_\Gamma(i\nu_n) = i2\pi \Gamma^2 N$. This corresponds to a Markovian 
approximation to the bath in which $f_\Gamma(\tau)=i2\gamma^2\delta(\tau)$. 
In the second limit, if one assumes that the matrix elements and density
of states are concentrated at zero frequency, one has $f_\Gamma(i\nu_n)=
2\gamma^2\delta_{\nu_n,0}$. Here, in the static limit, $f_\Gamma(\tau)=
2\gamma^2$ is \emph{real}. Applied to the self-energy, the static limit 
leads to 
\begin{eqnarray*}
\Sigma_{jj'}(i\nu_n) = \gamma^2 G_{jj'}(i\nu_n), 
\end{eqnarray*}
while, in the Markovian approximation,
\begin{eqnarray*}
\Sigma_{jj'}(\tau,\tau) = i\gamma^2 G_{jj'}(\tau,\tau),
\end{eqnarray*}
(similarly for the perturbation $\Gamma^{(2)}$).
In these two limiting cases, the saddle-point equations admit a 
straightforward analytical solution. Here we present a detailed analysis
for the static limit.

Noting that photon field in the gap equation (\ref{saddlepointF}) couples 
only to the diagonal elements of the Green function, $G_{jj}$, this suggests
a mean-field Ansatz in which the matrix elements off diagonal in the $j$
space are taken to be zero, while the only time dependence of the
field is associated with oscillation at the chemical potential, $\mu$.
In this case, the coupled equations (\ref{saddlepointF}), (\ref{Dyson}) 
can now be solved.

To simplify the algebra we employ a similar mathematical trick to that
used by Abrikosov and Gor'kov in their theory of gapless superconductors. 
Since the overall phase of a coherent state is arbitrary, we can choose the 
mean-field $\langle \psi \rangle$ to be real and present the total Green
function $G$ (\ref{Dyson}) in the same form as the zero-order Green function 
$G_0$,
\begin{equation}
G^{-1}_{jj}(i\nu_n)=-i\tilde{\nu}_{j,n}-(\tilde{\epsilon}_j-\mu/2)\sigma_3-
g\langle \tilde{\psi}_j \rangle \sigma_1,
\label{G2}
\end{equation}
using the frequency dependent, renormalised $\tilde{\nu}_{j,n}$,
$\tilde{\epsilon}_j$ and $\langle \tilde{\psi}_j \rangle$.  Comparing
(\ref{Dyson}) with (\ref{G2}) we obtain for both type 1 and type 2
decoherence three equations determining the renormalised frequency,
energy and coherent photon field
\begin{align}
\label{eq:renorm-omega}
\tilde{\nu}_{j,n} &=\nu_n-\gamma_{1,2}^2 \frac{\tilde{\nu}_{j,n}}
{\tilde{\nu}_{j,n}^2+(\tilde{\epsilon}_j-\mu/2)^2 + g^2\langle 
\tilde{\psi}_j\rangle^2},\\
\label{eq:renorm-epsilon}
\tilde{\epsilon}_j &= \epsilon_j 
+\gamma_{1,2}^2 \frac{\tilde{\epsilon}_j}
{\tilde{\nu}_{j,n}^2+(\tilde{\epsilon}_j-\mu/2)^2 + g^2\langle 
\tilde{\psi}_j\rangle^2},\\
\langle \tilde{\psi}_j \rangle &=\langle \psi \rangle
\pm \gamma_{1,2}^2 \frac{g_j\langle \tilde{\psi}_j \rangle}
{\tilde{\nu}_{j,n}^2+(\tilde{\epsilon}_j-\mu/2)^2 + g^2\langle 
\tilde{\psi}_j\rangle^2}.
\label{eq:renorm-field}
\end{align}
while the gap equation (\ref{saddlepointF}) takes the form
\begin{equation} 
\langle\psi\rangle=\frac{g}{2(\omega_c-\mu)} \sum_{j} {\rm Tr} \ G_{jj}
\sigma_1.
\label{saddlepointF2}
\end{equation}
The average coherent polarisation of the medium can be determined from
the off-diagonal part of the Green's function $G$ (\ref{G2})
\begin{eqnarray}
&&\frac{1}{N}\langle \sum_j a^{\dagger}_jb_j \rangle = \langle P \rangle
\nonumber\\ 
&&\qquad = -\beta^{-1}\sum_{\nu_n,j} \frac{g\langle \tilde{\psi}_j \rangle}
{\tilde{\nu}_{j,n}^2+(\tilde{\epsilon}_j-\mu/2)^2 + g^2\langle 
\tilde{\psi}_j\rangle^2}.
\label{pol}
\end{eqnarray}
Thus, at the mean-field level, we can see from Eqs.~(\ref{saddlepointF2}) 
and (\ref{pol}) that the two order parameters, the coherent polarisation 
and the coherent photon field, are coupled according to the relation
\begin{equation}
\langle \psi \rangle= -\frac{g}{\omega_c-\mu} \langle P \rangle.
\label{eq:field_pol}
\end{equation}
The ratio between the two order parameters is determined
by the chemical potential, which in the steady state can be calculated
from Eq.~(\ref{constrain}). 

The number of electronic excitations, refered to later as inversion, can
be obtained from the diagonal elements of the Green's function
\begin{eqnarray}
&&\frac{1}{2}\langle \sum_j(b^{\dagger}_jb_j-a^{\dagger}_ja_j) \rangle
\nonumber\\ 
&&\qquad\qquad =-\beta^{-1}\sum_{\nu_n,j} \frac{2(\tilde{\epsilon}_j-\mu/2)}
{\tilde{\nu}_{j,n}^2+(\tilde{\epsilon}_j-\mu/2)^2 +
g^2\langle \tilde{\psi}_j \rangle^2}.
\label{inv}
\end{eqnarray}

Using Eqs. (\ref{eq:renorm-omega})-(\ref{eq:renorm-field}) we can
determine the renormalised parameters $\tilde{\nu}_{j,n}$, 
$\tilde{\epsilon}_j$ and $\langle \tilde{\psi}_j \rangle$ as a functions 
of the bare parameters $\nu_n$, $\epsilon_j$, $\langle \psi \rangle$ and 
$\gamma$. In the case of type 1 decoherence processes we obtain 
\breakon
\begin{equation}
\label{eq:last-field}
\langle \tilde{\psi}_j \rangle=\frac{\langle \psi \rangle}{2}+
\frac{\sqrt{2}\langle \psi \rangle}{4E_j}\sqrt{E_j^2-4\gamma_1^2
-\nu^2_n+\sqrt{-16\gamma^2_1E_j^2+(E_j^2+4\gamma^2_1+\nu^2_n)^2}},
\end{equation}
\breakoff
and 
\begin{align}
\label{eq:last-omega}
\tilde{\nu}_{j,n}&=\frac{\nu_n \langle \tilde{\psi}_j\rangle}
{2\langle \tilde{\psi}_j \rangle - \langle \psi \rangle}, \\
\label{eq:last-epsilon}
\tilde{\epsilon}_j&=\frac{\epsilon_j \langle \tilde{\psi}_j
\rangle}{\langle\psi \rangle},  
\end{align} 
while for the type 2 decoherence processes we have
\breakon
\begin{multline}
\label{eq:last-field2}
\langle \tilde{\psi}_j \rangle=\frac{\langle \psi \rangle}{2}+ \\
\frac{\sqrt{2}\langle \psi \rangle}{4(\nu_n^2+g^2\langle \psi\rangle^2)}
\sqrt{E_j^2-2(\epsilon-\mu/2)^2+4\gamma_1^2-\nu^2_n+
\sqrt{16\gamma^2_2(\nu_n^2+g^2\langle \psi
\rangle^2)+(E_j^2-4\gamma^2_1+\nu^2_n)^2}}, 
\end{multline}
\breakoff
and 
\begin{align}
\label{eq:last-omega2}
\tilde{\nu}_{j,n}&=\frac{\nu_n \langle \tilde{\psi}_j\rangle}{\langle\psi 
\rangle}, \\   
\label{eq:last-epsilon2}
\tilde{\epsilon}_j&=\frac{\epsilon_j \langle \tilde{\psi}_j\rangle}
{2\langle \tilde{\psi}_j \rangle - \langle \psi \rangle},
\end{align}
where for both cases
\begin{equation}
\label{eq:Equasip}
E_j=\sqrt{(\epsilon_j-\mu/2)^2+g^2\langle \psi \rangle^2}.
\end{equation}
Substituting Eqs.~(\ref{eq:last-field})-(\ref{eq:last-epsilon}) or
(\ref{eq:last-field2})-(\ref{eq:last-epsilon2}) into Eq.~(\ref{pol}),
summing over the Matsubara frequencies, and using Eq.~(\ref{eq:field_pol}) 
we can determine the ground state coherent polarisation $\langle P \rangle$ 
and the coherent photon field $\langle \psi \rangle$ as functions of the 
system parameters $\epsilon$, $\omega_c$, decoherence parameter 
$\gamma_{1,2}$ and chemical potential $\mu$. The chemical potential can be 
then obtained from Eq.~(\ref{constrain}). The integrals over the Matsubara 
frequencies at zero temperature in (\ref{pol}) and (\ref{inv}) as well as the
determination of the chemical potential from (\ref{constrain}) have to
be performed numerically.

\subsection{Density of states}

The excitation spectrum, i.e the density of states, can be obtained
from the diagonal part of the Green function on the real frequency
axis. Considering the analytical continuation of $G_{jj}(i\nu_n) \to
\bar G_{jj}(\nu)$, where $\nu_n$ and $\nu$ are the Matsubara and the
real frequencies respectively, and thus using the usual substitution
$i\nu_n \to -\nu +i \delta$, we obtain the relationship between the
Green function $G_{jj}(i \nu_n)$ and the density of states $A(\nu)$
\begin{equation}
A(\nu)=\sum_j\lim_{\delta \to 0^+}{\rm Im} G_{jj}(-\nu+i\delta+\mu),
\end{equation}
which has the following form
\begin{equation}
\label{eq:spectrum}
A(\nu)=\sum_j {\rm Im} \frac{\tilde{\nu}_j+(\tilde{\epsilon}_j-\mu/2)}
{\tilde{\nu}_j^2+(\tilde{\epsilon}_j-\mu/2)^2 +
g^2\langle \tilde{\psi}_j \rangle^2}.
\end{equation}
Generally, $\tilde{\nu}_j$, $\tilde{\epsilon}_j$ and $\langle
\tilde{\psi}_j \rangle$ are functions of $\nu$, $\epsilon_j$, $\langle
\psi \rangle$ and $\gamma$ which can be determined from Eqs.
(\ref{eq:last-field})-(\ref{eq:last-epsilon}) or
(\ref{eq:last-field2})-(\ref{eq:last-epsilon2}) by the following
substitution $i\nu_n \to -\nu +i \delta$. It can be shown 
from Eq.~(\ref{eq:spectrum}) that the system of 
two-level oscillators with uniform energies, $\epsilon_j=\epsilon$, in
the presence of the type 1 processes has a gap, $\Delta$, in the
density of states of magnitude
\begin{equation}
\Delta=2\sqrt{(\epsilon-\mu/2)^2+g^2\langle \psi \rangle^2}-4\gamma_1.
\label{eq:gap}
\end{equation}
At very high excitation densities the gap is proportional to the
coherent field amplitude $\Delta \approx 2g\langle \psi \rangle
-4\gamma_1$. At very low excitation densities, when $\langle \psi
\rangle \to 0$, we recover conventional polaritons for
which the chemical potential is $2\epsilon-\mu \to g$ and the gap
$\Delta \to g-4\gamma_1$.

The major difference between this work and the AG theory
\cite{abrikosov-gorkov} is that the system studied here has two order
parameters connected through the chemical potential which needs to be
determined. We use a different form of the density of states for the
two-level oscillators than in their theory. Instead of a flat
distribution of energies from $-\infty$ to $+\infty$ used in the AG
method we first perform the calculations for the degenerate case where
all two-level oscillators have the same energy $\epsilon$ and then we
use a realistic Gaussian distribution of energies, present in the real
microcavities. To account for these differences we need to include the
additional, third equation for renormalised $\tilde{\epsilon}$ not
present in the original Abrikosov and Gor'kov method and the
constraint equation for $n_{ex}$. In the Abrikosov and Gor'kov theory
they consider free propagating electrons with momentum {\bf k} over
which all the summations are performed. In our model of the localised
two-level systems the summations are performed over the sites where
the two-level oscillators can be present. Dynamic impurities can
easily be included in our formalism.

To perform the calculations we rescale the coherent fields by
$\sqrt{N}$ and consequently the inversion and the number of
excitations by $N$ introducing the excitation density
$\rho_{ex}=n_{ex}/N$. In this terminology the minimum $\rho_{ex}=-0.5$
corresponds to no photons and no electronic excitations in the
system. The condition $\rho_{ex}=0.5$ in the absence of photons would
correspond to all two-level oscillators in excited states, thus to the
maximum inversion.

We calculate first the ground state coherent field $\langle \psi
\rangle$, the coherent polarisation $\langle P \rangle$, the inversion
and the chemical potential as functions of the decoherence strength
$\gamma$ and the excitation density $\rho_{ex}$ for different
distributions of excitonic energies. Then we study the excitation
spectrum of the system for different regimes. The ground state
properties and the excitation spectrum allow us to obtain a phase
diagram for different excitation densities and decoherence
strengths. We consider the influence of both type 1 and type 2
decoherence processes as well as inhomogeneous broadening of exciton
energies.

\section{The Ground State --- Coherent Fields}
\label{gs}

\subsection{Type 1 (Pair-Breaking) Decoherence Processes}
\label{type1}

To examine the ground state properties of the system in the presence
of the type 1 decoherence processes we study the mean value of the
annihilation operator of the field and the polarisation. This mean is
non-zero only in a coherent state.  Figure \ref{fig:phot} (upper
panel) shows the behaviour of the coherent part of the photon field
$\langle \psi \rangle$ as the decoherence strength $\gamma$ is changed
for different excitation densities $\rho_{ex}$ and different
inhomogeneous broadenings of exciton energies.

For small values of $\gamma/g$, up to some critical value
$\gamma_{C1}$, $\langle \psi\rangle$ is practically unchanged while
for $\gamma/g>\gamma_{C1}$ the coherent field is damped quite rapidly
with increasing decoherence. This critical value of the
decoherence strength, $\gamma_{C1}$ is proportional to $\rho_{ex}$,
suggesting that for higher excitation densities the system is more
resistant to dephasing. At low excitation densities, where
$\rho_{ex}<0$, there is a second critical value of the decoherence
strength, $\gamma_{C2}$, where both coherent fields are sharply damped
to zero. As the excitation density is increased, precisely at
$\rho_{ex}=0$, $\gamma_{C2}$ diverges and does not exist
for $\rho_{ex}>0$ --- coherent fields although reduced are never
completely suppressed.

The behaviour of the electronic inversion, which is a measure of the
excitonic density given by Eq.~(\ref{inv}), is presented in
Fig. \ref{fig:phot} (middle panel). In our terminology an inversion
of -0.5 corresponds to no excitons while an inversion equal to zero means 
that the excitonic system is half occupied (roughly 0.5 per Bohr radius in a
model where double occupation of excitonic sites is not
allowed). Inversion of 0.5 would then correspond to 1 exciton per Bohr
radius. In this region of the decoherence strength where $\langle \psi
\rangle$ is damped, the inversion increases. At low excitation
densities ($\rho_{ex}<0$) the inversion approaches $\rho_{ex}$ for
$\gamma/g=\gamma_{C2}$ and stays constant as $\gamma$ is further
increased. At high excitation densities ($\rho_{ex}>0$) the inversion
asymptotically approaches zero with increasing dephasing.  The maximum
value of electronic inversion, for any exciton density and decoherence
strength, is zero, which corresponds to a half filled excitonic
system. This is a consequence of our assumption of thermal equilibrium in the 
exciton-photon system.

The ratio of coherent polarisation to coherent field $\langle
P \rangle/\langle \psi \rangle$ is presented in Fig. \ref{fig:phot}
(lower panel). For an isolated system, where $\gamma=0$, this ratio
depends on the excitation density. The condensate becomes more
photon like as $\rho_{ex}$ is increased due to the phase space filling
effect. For finite $\gamma$, at a given excitation density, this ratio
decreases with increasing $\gamma$ meaning that the coherent
polarisation is more heavily damped than the coherent photon field by
the type 1 decoherence processes. At $\rho_{ex}<0$ this ratio becomes
undefined for $\gamma/g>\gamma_{C2}$ when both coherent fields vanish.

In order to study the system of realistic, inhomogeneously broadened
two-level oscillators we have replaced the summations over sites with
integrals over the energy distribution. We have assumed this
distribution to be a Gaussian with mean $\epsilon_0$ and variance
$\sigma $. Our results, presented as dashed and dashed-dotted lines in
Fig \ref{fig:phot} show that a Gaussian broadening of energies does
not make any qualitative difference to the degenerate case. The coherent
fields and the critical values of decoherence strength, $\gamma_{C1}$
and $\gamma_{C2}$ are, as expected, slightly smaller than in the
degenerate case but all the regimes are analogous.

\begin{figure}[htbp]
	\begin{center}
\includegraphics[width=8.8cm,angle=0]{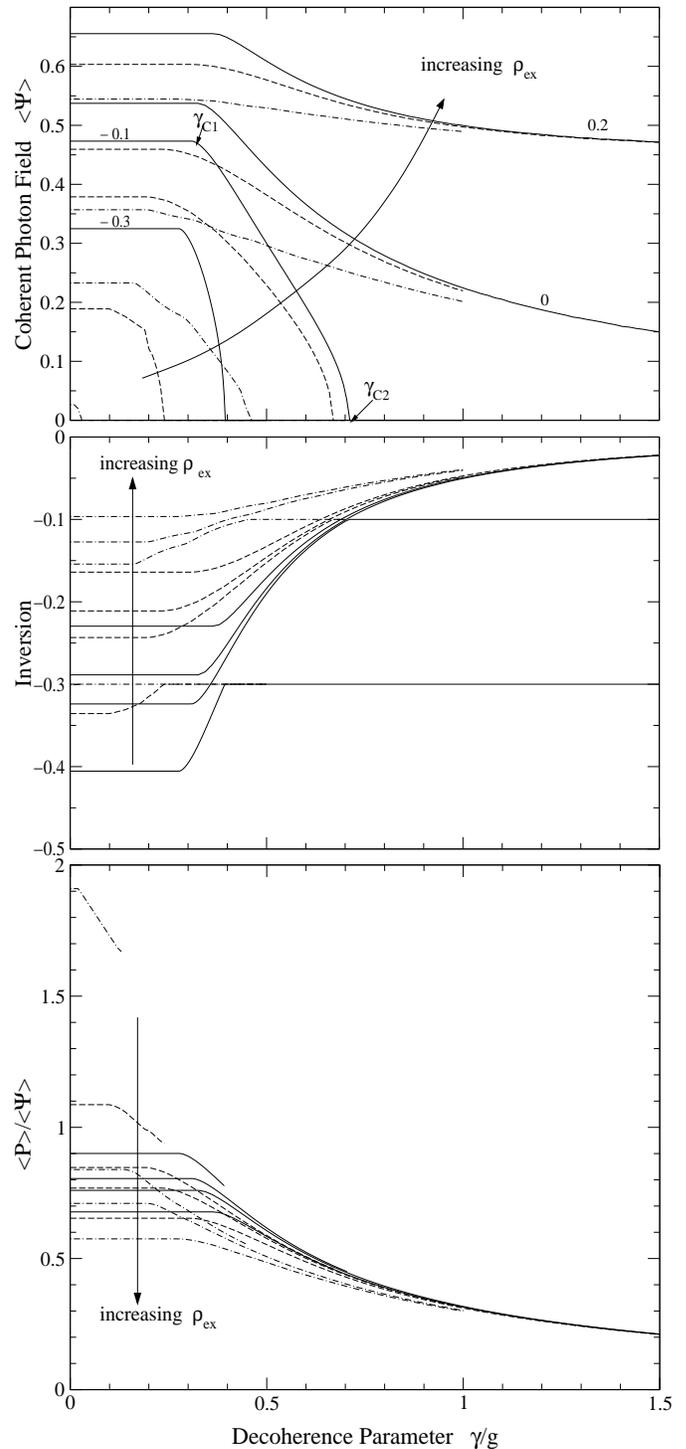}
	\end{center} 
	\caption{Coherent photon field $\langle \psi \rangle$ (upper
	panel), inversion (middle panel) and ratio between coherent
	photon field and coherent polarisation $\langle P
	\rangle/\langle \psi \rangle$ (lower panel) as functions of
	the pair-breaking decoherence strength, $\gamma/g$ for
	different excitation densities, $\rho_{ex}$ and variances of
	inhomogeneous broadening $\sigma=0$ (solid lines), $\sigma=0.5$
	(dashed lines) and $\sigma=1.0$ (dotted lines). }

\label{fig:phot}
\end{figure}

\subsection{Type 2 (Non-Pair-Breaking) Decoherence Processes}
\label{type2}

\begin{figure}[htbp]
	\begin{center} 
\includegraphics[width=8.8cm,angle=0]{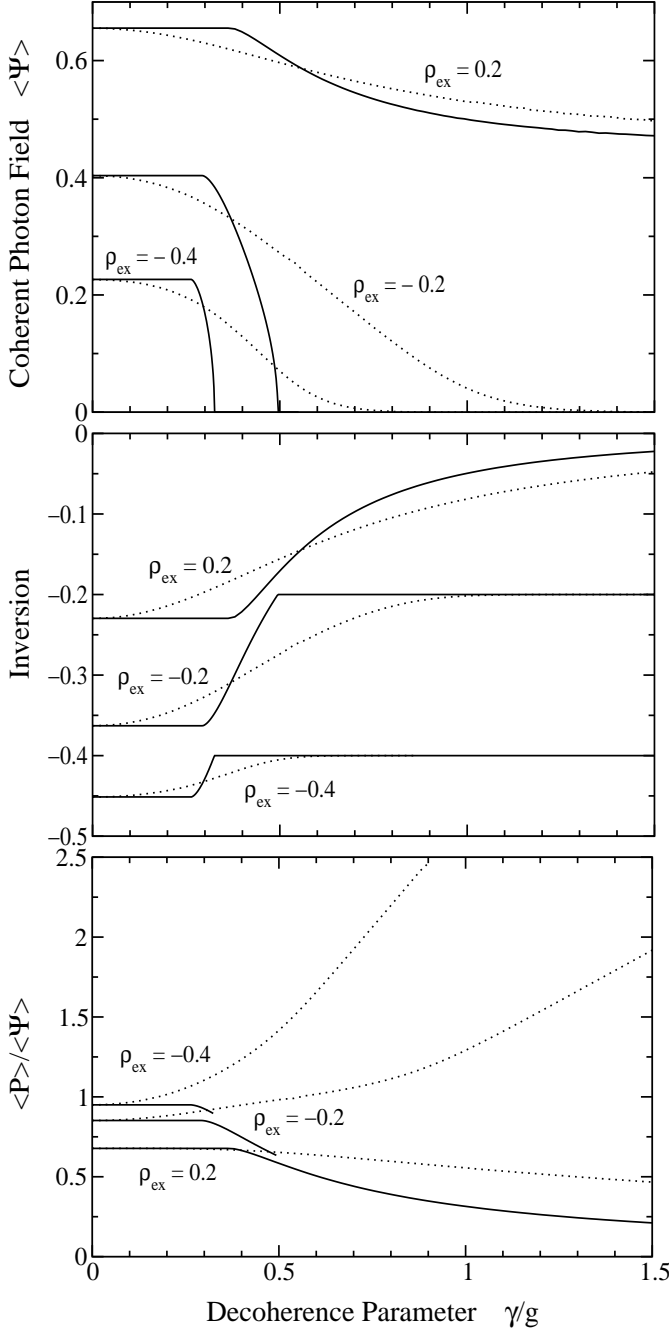}
	\end{center} 
	\caption{Comparison between the influence of a
	pair-breaking (solid line) and a non-pair-breaking (dotted
	line) decoherence processes on $\langle \psi \rangle$ (upper
	panel), inversion (middle panel) and $\langle P
	\rangle/\langle \psi \rangle$ (lower panel) for three different
	values of $\rho_{ex}$.}
\label{fig:photnonm}
\end{figure}

We now repeat the analysis for type 2 decoherence processes
(\ref{Hmag}), which act in an exactly the same way on the upper and
the lower levels of the two-level oscillator. Recall that such processes
mirror the effects of non-magnetic impurities in the superconductor.

In Fig. \ref{fig:photnonm} we present for comparison the coherent
photon field (upper panel), the inversion (middle panel), and the
ratio between the coherent polarisation and the coherent photon field
(lower panel) in the presence of the type 1 (solid line) and the type
2 (dotted line) decoherence processes for three excitation densities
$\rho_{ex}=-0.4$, $\rho_{ex}=-0.2$ and $\rho_{ex}=0.2$. It is evident
that the singular behaviour at $\gamma_{C1}$ and $\gamma_{C2}$
discussed for the type 1 decoherence processes is not present for the
case of the type 2 processes. With an increase of the decoherence
strength from zero the coherent fields are slightly reduced and they
slowly decrease, asymptotically approaching zero at low excitation
densities ($\rho_{ex}<0$) or a constant value at high densities
($\rho_{ex}>0$). Although both coherent fields are reduced the
behaviour of their ratio strongly depends on the excitation
density. We will show in Section \ref{nonmagnetic} that the type 2
processes give rise to the broadening of energies and then the
behaviour of the ratio $\langle P \rangle/\langle \psi \rangle$
depends on the position of the chemical potential with respect to the
energy distribution.  The type 2 processes can make the condensate
more photon or more exciton like depending on the parameters of the
system. The two different cases are presented in
Fig. \ref{fig:photnonm} (lower panel).

\section{Excitation Spectrum and the Phase Diagram}
\label{es}
To understand the behaviour of the coherent fields we study the
excitation spectrum of the system in different regimes.  The density
of states of the system with uniform energy distribution for six
different values of $\gamma$ at low excitation density
$\rho_{ex}=-0.4$ is presented in Fig. \ref{fig:green-0.4} while at
high excitation density $\rho_{ex}=-0.2$ in Fig. \ref{fig:green0.2}.

\subsection{Type 1 (Pair-Breaking) Decoherence Processes}
In the absence of decoherence (Fig. \ref{fig:green-0.4} a and
\ref{fig:green0.2} a) we have two sharp peaks at two quasi-particle
energies, $\pm E$, given by equation (\ref{eq:Equasip}) for the
degenerate case ($E_j=E$). As $\gamma$ increases these two peaks broaden,
which causes a decrease in the magnitude of the energy gap
(Fig. \ref{fig:green-0.4} b, c and \ref{fig:green0.2} b, c, d ). The
magnitude of the energy gap in the degenerate case is equal to
$2E-4\gamma$, which is given in more detail in equation
(\ref{eq:gap}). Finally, precisely at $\gamma_{C1}$ (shown in Fig.
\ref{fig:phot}), these two broadened peaks join together and the gap
closes (Fig. \ref{fig:green-0.4} d and \ref{fig:green0.2} e). When the
decoherence strength is increased further (Fig. \ref{fig:green-0.4} e)
these two peaks overlap more and the shape of the gapless density of
states changes. For $\gamma/g>\gamma_{C2}$ at low densities the
coherent fields are suppressed, thus Fig \ref{fig:green-0.4} f shows
the normal state density of states in the absence of coherence.  Finally, 
at high densities (Fig. \ref{fig:green0.2} f), as one would expect for a 
laser system, the coherent field is present without a sign of a gap.

\begin{figure}[htbp]
	\begin{center}
\includegraphics[width=8.8cm,angle=0]{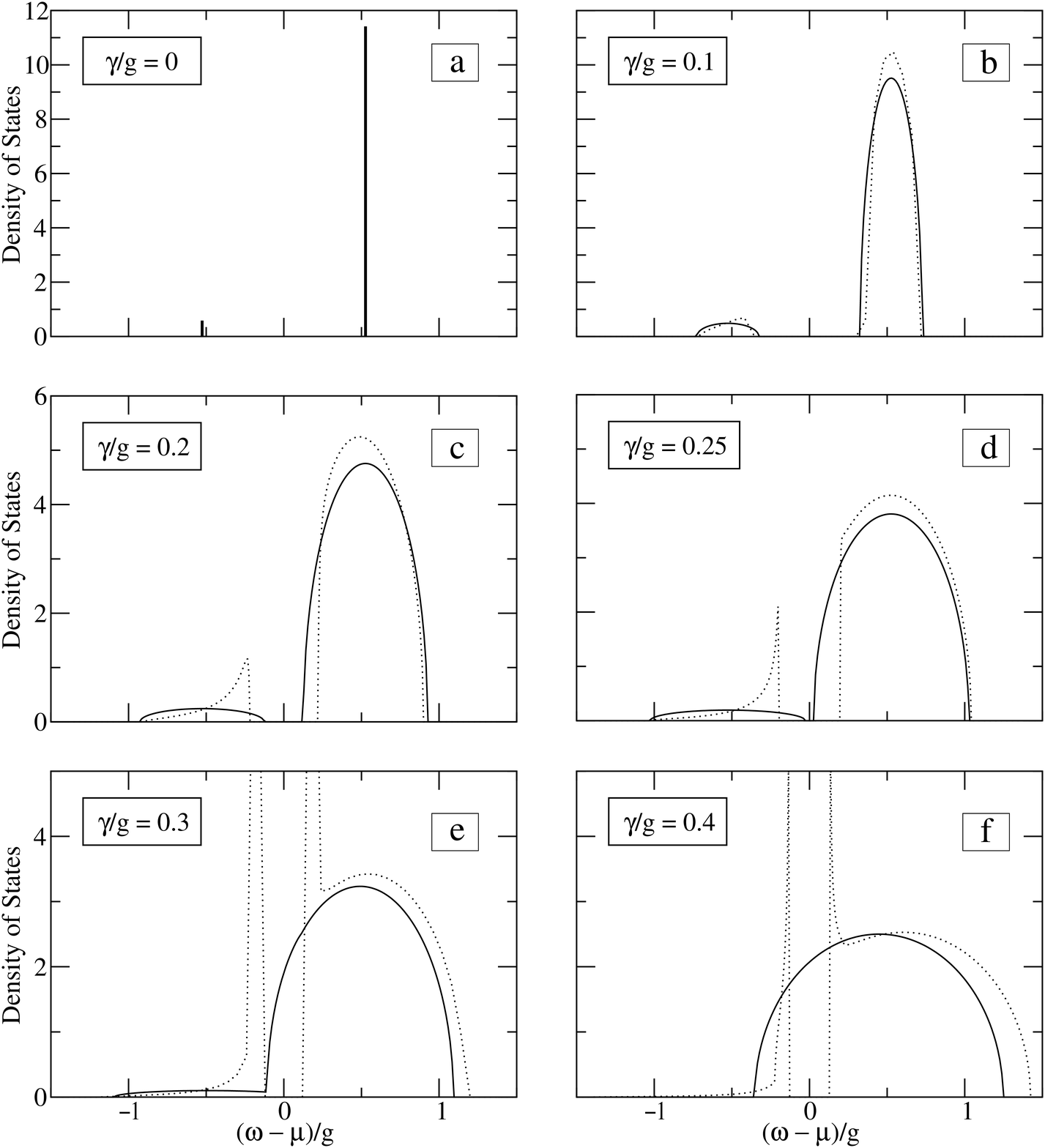} 
	\end{center}
	\caption{Density of states for $\rho_{ex}=-0.4$ and different
  	decoherence strengths, $\gamma/g$ for a pair-breaking (solid
  	line) and a non-pair-breaking (dotted line) decoherence
  	processes.}
\label{fig:green-0.4}
\end{figure}

Fig \ref{fig:greenb} shows the density of states for the system of the
inhomogeneously broadened two-level oscillators with standard
deviation $\sigma=0.5g$ for different values of the type 1 decoherence
strength at two different excitation densities.  The broadening of the
density of states and the suppression of the energy gap can be
observed as $\gamma$ is increased. The last curve in the Fig
\ref{fig:greenb} (left panel) shows the normal state density 
where the coherent fields are suppressed. In the Fig \ref{fig:greenb}
(right panel) where the coherent fields are present for all the values
of $\gamma$ the last curve has no sign of a gap.

There are clearly three different phases depending on the decoherence
strength, $\gamma$ (homogeneous broadening), inhomogeneous broadening,
$\sigma$ and the excitation density $\rho_{ex}$. In the Fig.
\ref{fig:phase} we present a phase diagram for the system.  The phase
boundaries are defined by $\gamma_{C1}$ and $\gamma_{C2}$ for
different values of $\rho_{ex}$ and $\sigma$.

Below the white surface, for small decoherence, we have a phase in
which both coherent fields and an energy gap in the density of states
are present. In this region coherent fields are protected by the
energy gap and remain practically unchanged while the energy gap
narrows as the decoherence is increased. At low densities within this
phase we have a BEC of polaritons with the electronic and photonic
parts comparable in size. At high densities we have a BCS-type of
condensate with photon component increasing with the excitation
density. Despite the predominantly photon-like character of this phase
at very high densities, the coherence of the medium is still large and
this phase can be distinguished from a laser by the presence of a gap in
the density of states (Fig \ref{fig:greenb} - right panel). The
crossover between high and low densities is a smooth evolution and
there are no rapid phase transitions.

\begin{figure}[htbp]
	\begin{center}
\includegraphics[width=8.8cm,angle=0]{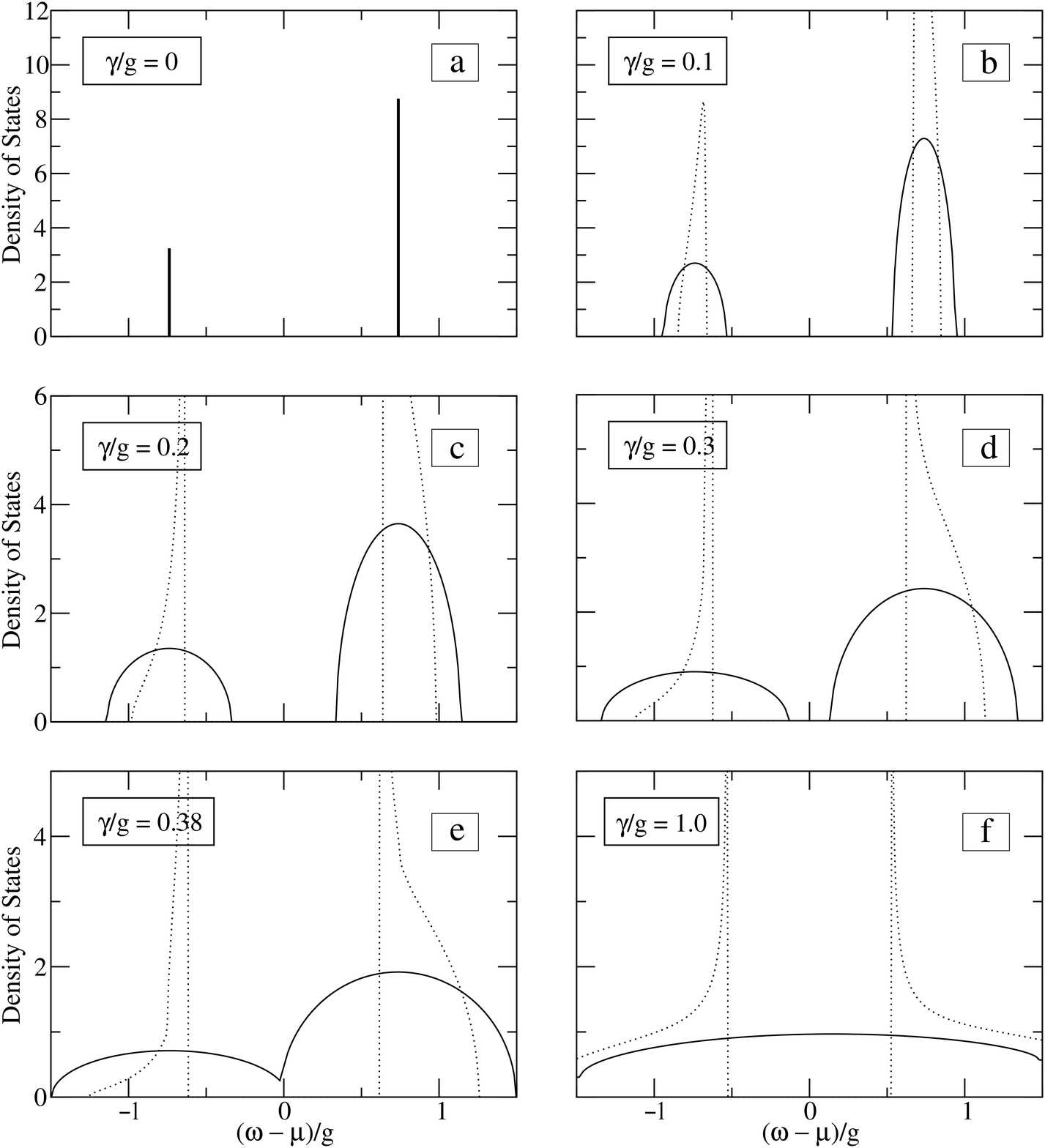}
	\end{center}
	\caption{Density of states for $\rho_{ex}=0.2$ and different
  	decoherence strengths, $\gamma/g$ for a pair-breaking (solid
  	line) and a non-pair-breaking (dotted line) decoherence
  	processes.}
\label{fig:green0.2}
\end{figure}

Inhomogeneous broadening has some influence on this phase mainly at low
excitation densities. The critical decoherence $\gamma_{C1}$, which
defines a boundary between gapped and gapless phases, decreases with an
increase in the inhomogeneous broadening $\sigma$. At high excitation
densities inhomogeneous broadening has very weak influence on the
system.   

Between the white and gray surfaces there is a phase where the
coherent fields are present without an energy gap in the density of
states. This phase exists for all values of the decoherence parameter
at high excitation densities $\rho_{ex} >0$. The coherent fields are no
longer protected by the gap and get reduced as the decoherence is
increased. The coherent polarisation is more heavily damped than the
coherent field.  Within this phase we have at low densities a gapless,
light-matter condensate, analogous to gapless superconductivity, whilst
at very high densities we have essentially the laser system.
Finally, above the dashed line there is a phase where the coherent
fields are completely suppressed.

\breakon
\begin{figure}[htbp]
	\begin{center}
\includegraphics[width=16.2cm,angle=0]{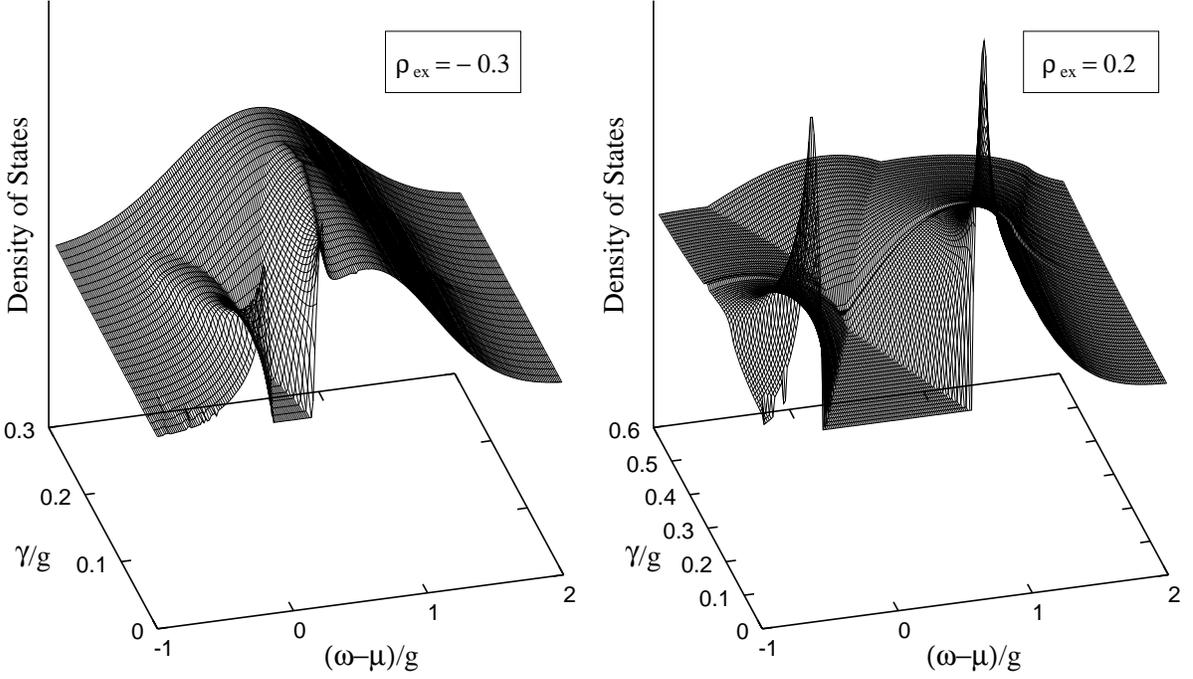}
	\end{center}
	\caption{Density of states for the Gaussian broaden case
	with $\sigma=0.5$ for different values of the pair-breaking
	decoherence strength at $\rho_{ex}=-0.3$ (left panel) and
	$\rho_{ex}=0.2$ (right panel). }
\label{fig:greenb}
\end{figure}    
\breakoff
\begin{figure}[htbp]
	\begin{center}
\includegraphics[width=8.8cm,angle=0]{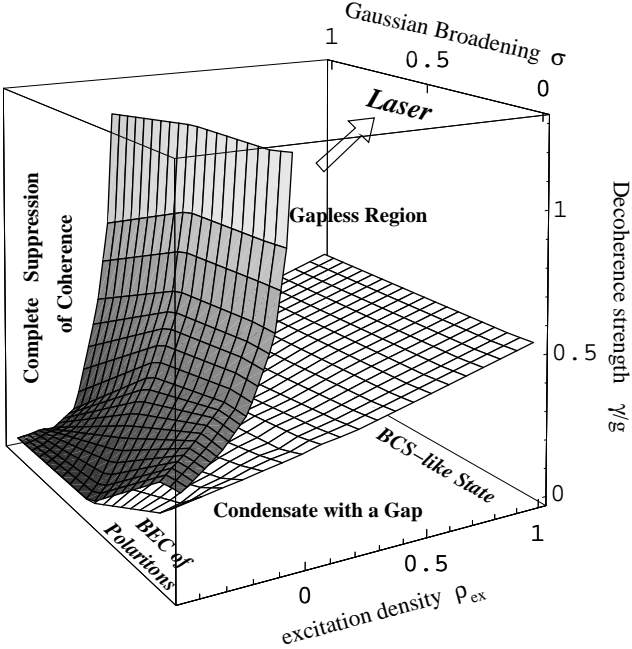}
	\end{center}
	\caption{Phase diagram. The phase boundaries $\gamma_{C1}$ and
	$\gamma_{C2}$ as marked in Fig \ref{fig:phot}.  }
\label{fig:phase}
\end{figure}

\subsection{Type 2 (Non-Pair-Breaking) Decoherence Processes}
\label{nonmagnetic} 

The density of states in the presence of type 2 processes is shown in
Figs \ref{fig:green-0.4} and \ref{fig:green0.2} (dotted lines). It can
be seen that, although the two quasiparticle peaks get very broad,
the gap is only slightly affected by the type 2 processes and is
always present. Even for much larger values of $\gamma$ than presented
in Figs \ref{fig:green-0.4} and \ref{fig:green0.2} the gap is not
suppressed. The gap in the density of states is present until the
coherent fields get completely suppressed. At high excitation
densities ($\rho_{ex}>0$) coherent fields are always present and thus
the density of states will have a gap for all values of the
decoherence strength.

The type 2 processes give a similar effect as the inhomogeneous
broadening of energy levels in the case of an isolated system (see Ref.
\cite{paul}). The difference is that the density of states in the
presence of the type 2 processes has sharp boundaries. This is a
result of the method which is used to perform the calculations. It can
be seen that if the bath operators for the type 2 processes,
$c_2$'s in expression (\ref{Hmag}), were just numbers, this
term could have been included into the second term in equation
(\ref{HS}) and would had just given a random $\epsilon_i$. In our
calculations the $c_2$'s are operators and we use a
self-consistent Born approximation, which is not exact, and does
not correctly reproduce the tail of the distribution. Thus, because 
of the way the
calculations are done, the density of states produced always has sharp
boundaries.

In the Abrikosov and Gor'kov theory, due to the flat density of states
used in the calculations, the non-magnetic impurities do not influence
the superconducting state at all. In the case of degenerate or realistic,
Gaussian distributed energies of the two level oscillators the type 2
processes have some quantitative influence on the coherent fields and
the gap but cannot cause any phase transitions.

\section{Crossover to Laser}
\label{cl}

The features of the second phase (between the white and the gray
surfaces) in Fig. \ref{fig:phase} at high $\rho_{ex}$ are essentially
the same as those of the laser system. The laser operates in the
regime of a very strong decoherence, comparable with the light-matter
interaction itself. In the presence of such a large decoherence laser
action can be observed only for a sufficiently large excitation
density.  The coherent polarisation in a laser system is much more
heavily damped than the photon field and the gap in the density of
states is not observed. Thus the laser is a regime of our system for
very large $\rho_{ex}$ and $\gamma$.

Laser theories, due to the approximations on which they are based, can
only be valid in a regime where the gap in the density of states is
suppressed and thus for a large decoherence. At the time when these
theories were proposed they were deemed sufficient as most lasers
operate in such a regime. Miniaturisation and improvements in the
quality of optical cavities in recent years can lead to a large
suppression of decoherence in laser media. For small decoherence and
very small pumping in comparison to decay processes, when $\rho_{ex}$
is much smaller then $0$, the laser theories would predict a lack of
coherence while the real ground state of the system would be a more
matter-like condensate. Thus an extension of laser theories to account
for the gap in the excitation spectrum and coherence in a media is
necessary. In our theory the laser emerges from the polariton
condensate at high densities when the gap in the density of states
closes for large decoherence and thus is analogous to a gapless
superconductor.

When both the gap and the coherence in a media are taken into account,
in contrast to a traditional laser, the coherent photon field can be
present without the population inversion in the media. The polariton
condensate is thus an example of a laser without inversion.

However, it has to be pointed out that there is no formal distinction
between a laser and a Bose condensate of polaritons.  In the laser,
coherence in the medium (manifested by the coherent polarisation)
although small, is not completely suppressed so the laser can be seen
as a gapless condensate with a more photon-like character. One of the
possible distinctions between BEC and laser could be an existence of
an energy gap in the excitation spectrum.

\section{The gap}
\label{gap}
\begin{figure}[htbp]
	\begin{center}
 \includegraphics[width=8.8cm,angle=0]{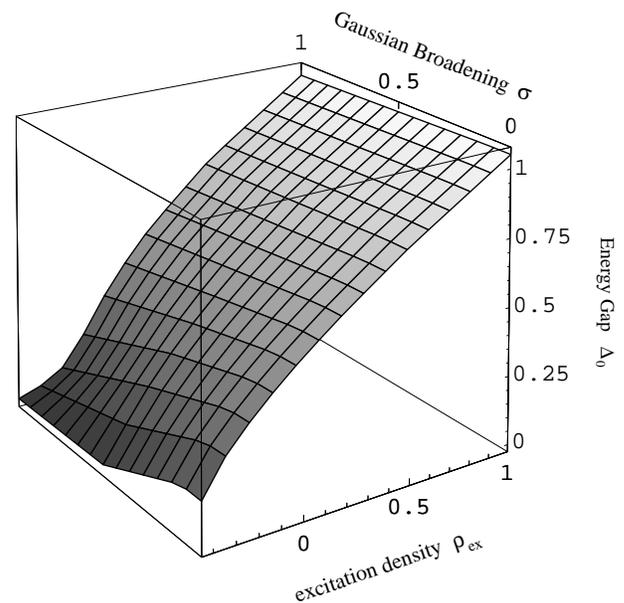}	
	\end{center}
	\caption{Energy gap $\Delta_0$ in the units of g in the
	absence of decoherence as a function of excitation density
	$\rho_{ex}$, and Gaussian broadening, $\sigma$.}
\label{fig:del0}
\end{figure}

In the degenerate case the magnitude of the gap in the density of states
is given by (\ref{eq:gap})  If we consider an inhomogeneous broadening
of excitonic energies with a Gaussian distribution then there will always
be some oscillators with energies which make the term $(\epsilon -
\mu/2)^2$ vanish and thus, strictly, the magnitude of the gap would be
$2g\langle \psi \rangle-4\gamma_1$ at all excitation densities. This
means that the gap would vanish at $\langle \psi \rangle \to 0$ in
contrast to degenerate case in which its magnitude would be $g - 4
\gamma_1$. However, the contribution of these states, arising only from
the tails of the Gaussian distribution, would be very small
essentially leading to a soft gap of the magnitude of $g - 4
\gamma_1$ at very low densities as in degenerate case. 

We call $\Delta_0$ the gap in the absence of decoherence when
$\gamma_1=0$. Omitting the case when $n_{exc}=0$, discussed above, we
calculate numerically the magnitude of the gap, in the units of the
dipole coupling $g$, for a wide range of excitation densities and
inhomogeneous broadening $\sigma>0$. $\Delta_0$ very much depends on
the excitation density. The Gaussian broadening of energies decreases
the gap at low densities but has almost no influence at high
excitation densities. In the presence of the decoherence (homogeneous
broadening) the actual gap $\Delta$ would be
\begin{equation} 
\Delta = \Delta_0-4\gamma_1
\label{totgap}
\end{equation} 
Fig. \ref{fig:del0} and equation (\ref{totgap}) can be used to
estimate the magnitude of the gap in particular experimental systems
in which the values of broadenings, density and the dipole coupling is
known.  Some estimates of the magnitude of the energy gap in
particular materials and conditions were reported in a brief report
\cite{us}.

\section{Applicability of the model and the method}
\label{aplic}

As already stated in Section \ref{model}, the model of two-level
systems interacting only via cavity photon field gives a reliable
description of  tightly bound, Frenkel-type excitons localised
by disorder or bound on impurities, molecular excitons in organic
materials, atoms in the solid state or Josephson junctions arrays in
microwave microcavities. In the classical limit $lim_{N \rightarrow
\infty} (\langle n_{ex}\rangle /N) = \rho \rightarrow const.$ it has 
an exact solution of the mean-field form
\begin{equation}
\label{varwfn3}
|\lambda,u,v\rangle = e^{\lambda \psi^\dagger }
\prod_{j} ( v_{j} b_j^{\dagger} + u_{j} a_j^{\dagger} )|0\rangle.
\end{equation}
However, the case of Wannier excitons propagating in the sample,
present in some clean semiconductors (for example GaAs quantum wells), is
conceptually not very different. The sums over sites in the microscopic
Hamiltonian would have to be replaced with sums over momentum
states and Coulomb interactions between electrons and holes would have
to be included. Such a model without photons was studied by Keldysh
and Kopaev \cite{keldysh1} and for coherently driven semiconductor by
Schmitt-Rink et al. \cite{schmitt}. If we treat the Coulomb
interaction on the mean-field level then the natural extension of the
Keldysh mean-field wave-function to account for photons will be
\begin{equation} 
\label{varwfn2} 
|\Psi_0> = e^{\lambda 
\psi_0^\dagger}\prod_{\vec k} [ u_{\vec k} + v_{\vec k} b^\dagger_{k}
a_{k} ]|0\rangle \;\;\;,  
\end{equation}
which is of analogous form to that for the two-level systems case
(\ref{varwfn3}). Then the influence of the various decoherence
processes can be included in exactly the same way as described in
Section \ref{model}.

At the level of mean-filed, we do not expect that the change of the 
model from that of two-level oscillators to free electron and holes with a 
Coulomb interaction would cause any dramatic differences. It was
shown within the mean-field techniques of Keldysh
\cite{keldysh1,keldysh2,nozieresex1} that the Coulomb interaction has
a pairing effect and leads to a formation of a coherent excitonic
insulator phase and this would only be enhanced by the dipole
coupling.

At high excitation densities (large photon fields) the dipole
interaction between excitons and photons is the dominant
interaction. In the case of a driven excitonic system Schmitt-Rink
{\it at al} \cite{schmitt} pointed out that for a very large pumping,
and thus high excitation densities, the Coulomb interaction is just a
very minor correction to the dominant dipole coupling. They also state
that, in the absence of the Coulomb interaction, their results for
propagating electrons and holes are equivalent to those obtained for
an ensemble of independent two-level oscillators as optical
transitions with different {\bf k} decouple.

The Coulomb interaction would be more important at low densities.
However, in this case, the dominant contribution arising from the 
interaction between an electron and hole within the same
exciton, is taken into account in model studied in this work. All
other Coulomb interactions are much weaker and, as with the dipole
coupling, favour condensation. 

Although we do not expect qualitative differences, it would be
interesting, especially at low excitation densities, to perform similar
calculations in the momentum representation with the Coulomb
interaction. This would allow one to study the relative influence of the
two independent pairing mechanisms, namely the dipole coupling and
Coulomb interactions, on the condensate.

Summarising, our method of including and studying the decoherence
effects is general and can be applied in the same way as in this work
to propagating excitons as well as to coherently driven
condensates. The coherent photon field in the driven system is an
external, fixed, parameter and not a self-consistent field satisfying
equation (\ref{saddlepointF2}). In contrast the gap in the density of
states, proportional to the coherent field amplitude, is present
exactly as for an equilibrium condensate.  The model can be applied to
obtain a qualitative description of the physical behaviour even for
propagating or weakly bound excitons. It would give similar
predictions to the model based on propagating electrons and holes with
Coulomb interactions treated within the mean-field approximation. It
does not, however, include screening and other non mean-fields
effects.

If the microscopic origins of the decoherence processes are known for
a particular experimentally studied system, the theory could be
extended to include this information. The detailed account of the
coupling constants and the density of states for the environment can
be easily included within this framework. In this work we have assumed
a static bath or Markovian (Ohmic) bath but in general the bath
propagators are frequency dependent and thus $\langle \psi \rangle$
would depend on frequency as in the Eliashberg theory of gapless
superconductors \cite{eliash,scalapino}. The phenomenological
constants $\gamma_1$ and $\gamma_2$ can thus be, in principle,
obtained from this microscopic calculation for a particular system.

The results presented in this work are performed at zero temperature
for which the summations over the Matsubara frequencies in
(\ref{saddlepointF2}), (\ref{pol}) and (\ref{inv}) become integrals
and are performed numerically. The extension to finite temperatures is
straightforward. At finite temperatures the summations over a discrete
$\nu_n$ can be performed using the standard techniques
\cite{green} and thus finite temperatures could be easily studied.

A much more important extension than the finite temperature case is
the problem of non-equilibrium systems. For a system with very strong
pumping and decay processes the thermal distribution of the relevant
quasiparticles cannot be assumed. The influence of non-equilibrium
distributions on the polariton condensate will be published elsewhere.

\section{Remarks on Experimental Realisation of BEC of polaritons}
\label{exp}

While stimulated scattering into the lower polariton branch (reminiscent
of Bose statistics) was clearly observed and reported by several groups
\cite{dang,senbloch,boef,alex,bulkboser,deng,baumberg,angleboser,angle-cw-stimscat,nature},
there is still no evidence of excitonic coherence, and therefore BEC of
polaritons. The experiments that have been performed fall into two 
categories. In the first, 
polaritons are pumped coherently, with a laser having an angle of
incidence chosen at a ``magic angle'' close to the bottom of the lower
polariton branch. In the second set of experiments the pumping is
incoherent. The laser pumping is tuned to be well above the ground
state so that polaritons have to undergo several inelastic scattering
events before reaching the ground state and thus loosing the initial
coherence of the pump.

Although not yet measured, the potential coherence of polaritons in the
experiments of the first category would be inherited from the pump and
the behaviour could be explained in terms of parametric
oscillation. Experiments of the second category could be potential
candidates for spontaneous BEC of polaritons but evidence for it are
very sparse. The coherence of the photon field
emitted at polariton frequency has recently been measured in one
experiment \cite{deng} but this alone does not prove that there is a
coherence in the excitonic part. A laser is an example where coherent
photon field is generated without large excitonic coherence. A more
direct evidence of excitonic coherence is necessary. Large excitonic
coherence would map to a large gap in the excitation spectrum which
should be seen in the the incoherent luminescence.

Experiments on cavity polaritons which report the stimulated
scattering into the lower polariton branch are performed at low
densities to avoid the fermionic phase space filling effect, so that
two clear polariton peaks could be seen.  For such densities the gap
in the excitation spectrum would be very small and easily suppressed
by the decoherence processes in the sample. Indeed such a gap has
never been observed in the photoluminescence spectrum. The attempt to
increase the density of polaritons in these experiments results not in
the formation of the condensate but in a switching into the
weak-coupling regime and lasing.

This is not surprising since the increase in the density of polaritons
in these experiments is obtained by increasing the pumping of
excitons. Incoherent pumping is a pair-breaking decoherence process
and thus the increase in the pumping intensity results in the increase
of not only the density of excitons but also decoherent scattering. As
shown by Eastham and Littlewood \cite{paul}, the fermionic structure
of excitons does not prevent condensation, even at very high
densities. Thus, in the current experiments, we suggest that it is not
a phase-space filling effect which lead to a laser as the pump
intensity is increased, but the increase in the decoherence
strength. If the polariton condensate is ever to be observed an
increase in density without an increase in decoherence is necessary.

Localised and tightly bound excitons, such as excitons in disordered
quantum wells or Frenkel-type excitons in organic compands, seem to be
better candidates for observing a polariton condensate than the
high-quality GaAs quantum wells with weakly bound, delocalised
excitons. Static disorder is not a pair-breaking effect and would have
a weaker influence on the condensate than screening and ionisation in
the case of delocalised or weakly bound excitons.  Additionally the
polariton splitting reported in organic materials is as large as 80
meV \cite{organicpol1,organicpol2} in comparison to the upper bound of
20 meV in GaAs. Another very good candidate for observation of
effects described in this work would be atoms in solid state, glass
spheres, dilute atomic gases in which the dephasing is particularly
weak in comparison to the dipole coupling.

\section{Summary}
\label{sum}

We have studied the equilibrium Bose condensation of cavity polaritons
in the presence of decoherence. It has been shown \cite{us0} that the
widely used 
Langevin equations with the constant, frequency independent decay
rates are not valid for systems with large polarisation in which the
gap in the density of states is present. In the regime of weak
decoherence these processes have to be treated
self-consistently, in such a way that the gap in the density of states is
taken into account. We have proposed a self-consistent method
analogous to the Abrikosov and Gor'kov theory of gapless
superconductivity \cite{abrikosov-gorkov} which allow us to study all
regimes of the system as the decoherence is changed from zero to large
values and for a wide range of excitation densities.

We have found that at small decoherence the polariton condensate is
protected by the energy gap in the excitation spectrum. The gap is
proportional to the coherent field amplitude and thus the excitation
density, so the condensate, unlike the excitonic insulator
\cite{nozieresex1}, is more robust at high densities. This gap gets
smaller and eventually is completely suppressed as the decoherence is
increased.  We have shown that there is a regime, analogous to the
gapless superconductor, when the coherent fields are present without
an energy gap. This regime, at very high excitation densities, has
most of the features of a photon laser.

We have studied the influence of two different types of processes,
both pair-breaking and non-pair-breaking ones as well as inhomogeneous
broadening of the energy levels. We have shown that only the
pair-breaking processes can lead to phase transitions.
Non-pair-breaking processes and inhomogeneous broadening of energies
can quantitatively change the behaviour of the system but cannot
prevent condensation.

We have studied a general form of the interactions, introducing
pair-breaking and non-pair-breaking processes and thus our theory is
applicable to a wide range of systems. For more detailed results, the
particular origin of the interactions and thus the particular density
of states for the baths have to be taken into account.

We have studied the whole phase diagram of the system given by the
Hamiltonian (\ref{H}) for different values of the decoherence
strength, inhomogeneous broadening and the excitation densities and
established the crossover between an isolated polariton condensate and
a photon laser as the decoherence strength is increased.

Our results suggest that, in contrast to traditional lasers, 
coherent light can be generated without a population inversion. This
work generalised the existing laser theories to include the gap in the
excitation spectrum caused by the coherence in a media in the low
decoherence regime.

\section{Acknowledgement}
We would like to thanks F. M. Marchetti and P. R. Eastham for helpful
discussions concerning mathematical techniques. 
MHS acknowledge the support of research fellowship from Gonville and
Caius College, Cambridge.

\end{multicols}


\begin{references}

\bibitem{abrikosov-gorkov}
A.~A.~Abrikosov, L.~P.~Gor'kov, JETP {\bf 12}, 1243 (1960)

\bibitem{atompol} 
M.~G. Raizen, R.~J. Thompson, R.~J. Breccha, H.~J. Kimble, and H.~J.  
Carmichael, {\em Phys. Rev. Lett.} {\bf 63}, 240 (1989).

\bibitem{cavpol}
C. Weisbuch, M. Nishioka, A.Ishikawa, and Y.Arakawa, {\em
Phys. Rev. Lett.} {\bf 69}, 3314 (1992).

\bibitem{bulkcavpol} 
A. Tedicucci, Y. Chen, V. Pellegrini, M. Borger, L. Sorba, F. Beltram,
and F. Bassani, {\em Phys. Rev. Lett.} {\bf 75}, 3906 (1995).

\bibitem{organicpol1} 
D.~G. Lidzey, D.~D.~C. Bradley, M.~S. Skolnick, T.~Virgili, S.~Walker,
and D.~M. Whittaker, {\em Nature} {\bf 395}, 53 (1998).

\bibitem{organicpol2}
D.~G. Lidzey, D.~D.~C. Bradley, T.~Virgili, A.~Armitage,
M.~S. Skolnick, and S.~Walker, {\em Phys. Rev. Lett.} {\bf 82}, 3316 (1999).

\bibitem{chargedpol}
R. Rapaport, R. Harel, E. Cohen, A. Ron, and E. Linder,
{\em Phys. Rev. Lett.} {\bf 84}, 1607 (2000).

\bibitem{JJ}
P. Barbara, A. B. Cawthorne, S. V. Shitov, and C. J. Lobb, 
{ \em Phys. Rev. Lett.} {\bf 82}, 1963 (1999).

\bibitem{paul} 
P.~R. Eastham, P.~B. Littlewood, {\em Solid State Commun.} {\bf
116}, 357 (2000), { \em Phys. Rev. B} {\bf 64}, 235101 (2001). 


\bibitem{laser} 
For the quantum theory of the laser developed by
Haken, Risken, Lax, Louisell, Scully and Lamb see M.~O. Scully and
M.~S. Zubairy, {\em Quantum Optics} (Cambridge University Press,
Cambridge, U.K., 1997).


\bibitem{haken}
H.~Haken, {\em Rev. Mod. Phys.} {\bf 47}, 67 (1975); 
{\em Laser Theory}, Springer-Verlag 1984.

\bibitem{us0} 

M.~H.~Szymanska, P.~B. Littlewood, unpublished.  In
(\ref{maxwell-bloch}) the life-time for a single two-level oscillator
is generalised to many two-level systems without taking into account
any collective effects. It can be shown that if we couple every
two-level system to its own distinct bath we recover Equation
(\ref{maxwell-bloch}). Such a method and so (\ref{maxwell-bloch}) is
unphysical if the interactions between two-level system generated for
example by the photon field is substantial. The environment needs to
be coupled to the whole ensemble of two-level oscillators and averaged
out in the selfconsistent way.


\bibitem{wrong}
K. Hannewald, S Glutsch and F. Bechstedt, {\em
J. Phys. Condens. Matter} {\bf 13}, 275 (2001)  

\bibitem{keldysh1}
L. V. Keldysh and Y. V. Kopaev, { \em Fiz. Tverd. Tela} {\bf 6}, 2791
(1964), [Sov. Phys. Solid State {\bf 6}, 2219, (1965)].

\bibitem{keldysh2}
L. V. Keldysh and A. N. Kozlov, {\em Zh. Eksp. Teor. Fiz.}  
{\bf 54}, 978 (1968), [Sov. Phys. JETP {\bf 27}, 521 (1968)].

\bibitem{zittart}
J.~Zittartz, {\em Phys. Rev.} {\bf 164}, 575 (1967)

\bibitem{us}
M.~H.~Szymanska,  P.~B. Littlewood,
{ \em Solid State Commun.} {\bf 124}, 103 (2002)

\bibitem{schmitt}
S. Schmitt-Rink, D.~S. Chemla, and H. Haug, { \em Phys. Rev. B} {\bf
37}, 941 (1988).

\bibitem{nozieresex1}
C. Comte and P. Nozi{\`{e}}res, {\em J. Phys.} (Paris) {\bf 43}, 1069
(1982).

\bibitem{eliash}
G. M. Eliashberg, {\em Zh. Eksperim. i Teor. Fiz.} {\bf 38}, 966
(1960), [Sov. Phys. JETP {\bf 11}, 696 (1960)].

\bibitem{scalapino}
D.~J.~Scalapino, in {\em Superconductivity}, edited by R.~D. Parks,
volume~1, chapter~10, pages 449--560, Marcel Dekker, Inc., New York,
1969.

\bibitem{green} 
G. Rickayzen, {\em Green's Functions and Condensed Matter}
(Academic Press Inc., London, 1987).


\bibitem{dang}
L.~S. Dang, D.~Heger, R.~Andr{\'{e}}, F.~B{\oe}uf, and R.~Romestain,
{ \em Phys. Rev. Lett.} {\bf 81}, 3920 (1998).

\bibitem{senbloch}
P. Senellart and J. Bloch, { \em Phys. Rev. Lett.} {\bf 82},  1233  (1999).

\bibitem{boef} F. Boef, R. Andre, R. Romestain, L. S. Dang, {\em Phys. Rev. B}
{\bf 62}, R2279 (2000).

\bibitem{alex} A. Alexandrou, G. Bianchi, E. Peronne, B. Halle, F. Boeuf,
R. Andre, R. Romestain, L. S. Dang, {\em Phys. Rev. B}
{\bf 64}, 233318 (2001).

\bibitem{bulkboser}
V. Pellegrini, R. Colombelli, L. Sorba, and F. Beltram, { \em Phys. Rev. B}
{\bf 59}, 10059 (1999).

\bibitem{deng} H. Deng, G. Weihs, C. Santori, J. Bloch, Y. Yamamoto, { \em
Science} {\bf 298}, 199 (2002).

\bibitem{baumberg} 
J. J. Baumberg, P. G. Savvidis, R. M. Stevenson, A. I. Tartakovskii,
M. S. Skolnick, D. M. Whittaker, and J. S. Roberts, { \em
Phys. Rev. B} {\bf 62}, R16247 (2000).

\bibitem{angleboser}
P.~G. Savvidis, J.~J. Baumberg, R.~M. Stevenson, M.~S. Skolnick, D.~M.
Whittaker, and J.~S. Roberts, { \em Phys. Rev. Lett.} {\bf 84}, 1547
(2000).

\bibitem{angle-cw-stimscat} 
R.~M. Stevenson, V.~N. Astratov, M.~S. Skolnick, D.~M. Whittaker,
E.~Emam-Ismail, A.~I. Tartakovskii, P.~G. Savvidis, J.~J. Baumberg,
and J.~S.  Roberts, {\em Phys. Rev. Lett.} {\bf 85}, 3680 (2000).

\bibitem{nature} 
M. Saba, C. Ciuti, J. Bloch, V. Thierry-Mieg, R. Andre, Le Si Dang,
S. Kundermann, A. Mura, G. Bongiovanni, J. L. Staehli, B. Deveaud,
{\em Nature} {\bf 414}, 731 (2001)

\bibitem{atoms}
C. Liu, Z. Dutton, C. H. Behroozi, L. V. Hau, {\em Nature} {\bf 409},
490 (2001) 




\end{references}
\end{document}